\documentclass[aip,preprint,12pt]{revtex4-1}
\usepackage{amsmath}
\usepackage{amssymb}
\usepackage[latin1]{inputenc}
\usepackage[english]{babel}
\usepackage{fancyhdr}
\usepackage[dvips]{graphicx}
\begin{document}
\def\ep#1{\left(#1\right)}
\def\ec#1{\left[#1\right]}
\def\ea#1{\left\{#1\right\}}
\title{\bf A nonlinear Schr\"odinger equation for water waves on finite depth with constant vorticity}
\author{R. Thomas}
\email{thomas@irphe.univ-mrs.fr}
\affiliation{Institut de Recherche sur les Ph\'enom\`enes hors \'Equilibre,\\49, rue F. Joliot Curie B.P. 146 13384 Marseille Cedex 13 France}
\author{C. Kharif}
\affiliation{\'Ecole Centrale Marseille,\\38, rue Fr\'ed\'eric Joliot-Curie 13451 MARSEILLE Cedex 20}
\author{M. Manna}
\affiliation{Universit\'e de Montpellier II,\\Place Eug\`ene Bataillon 34095 MONTPELLIER}

\date{\today}

\begin{abstract}
A nonlinear Schr\"odinger equation for the envelope of two dimensional surface water 
waves on finite depth with non zero constant vorticity is derived, and the influence 
of this constant vorticity on the well known stability properties of weakly nonlinear 
wave packets is studied. It is demonstrated that vorticity modifies significantly the 
modulational instability properties of weakly nonlinear plane waves, namely the growth 
rate and bandwidth. 
\newline
At third order we have shown the importance of the coupling between the mean
flow induced by the modulation and the vorticity.
\newline
Furthermore, it is shown that these plane wave solutions 
may be linearly stable to modulational instability for an opposite shear current independently 
of the dimensionless parameter $kh$, where $k$ and $h$ are the carrier wavenumber and depth 
respectively.  
\end{abstract}


\maketitle

\def\abs#1{{\left\vert#1\right\vert}}
\def\oOmega{\overline\Omega}
\section{Introduction}\label{intro}
Generally, in coastal and ocean waters, the velocity profiles are typically established by bottom friction 
and by surface wind stress and so are varying with depth. 
Currents generate shear at the bed of the sea or of a river. 
For example ebb and flood currents due to the tide may have an important effect on waves and wave packets. 
In any region where the wind is blowing there is a surface drift of the water and water waves are particularly sensitive 
to the velocity in the surface layer.

Surface water waves propagating steadily on a rotational current have been studied by many authors. 
Among them, one can cite Tsao \cite{tsao}, Dalrymple \cite{dalry}, Brevik \cite{brevik}, Simmen \& Safmann \cite{simsaf}, 
Teles da Silva \& Peregrine \cite{silvaper}, Kishida \& Sobey \cite{kishiso}, Pak \& Chow \cite{pakchow}, 
Constantin \cite{constantin}, etc. For a general description of the problem of waves on current, 
the reader is referred to reviews by Peregrine \cite{peregrine}, Jonsson \cite{jonsson} and Thomas \& Klopman \cite{thomklop}. 
On the contrary, the modulational instability or the Benjamin-Feir instability of progressive waves in the presence 
of vorticity has been poorly investigated. Using the method of multiple scales Johnson \cite{johnson} examined 
the slow modulation of a harmonic wave moving over the surface of a two dimensional flow of arbitary vorticity. 
He derived a nonlinear Schr\"{o}dinger equation (NLS equation) with coefficients that depend, in a complicated way, 
on the shear and gave the condition of linear stability of the nonlinear plane wave solution by writing that the product 
of the dispersive and nonlinear coefficients of the NLS equation is negative. He did not develop a detailed stability 
analysis as a function of the vorticity and depth. Oikawa, Chow \& Benney \cite{oikawacb} considered the instability 
properties of weakly nonlinear wave packets to three dimensional disturbances in the presence of shear. 
Their system of equations reduces to the familiar NLS equation when confining the evolution to be purely two dimensional. 
They illustrated their stability analysis for the case of a linear shear. 
Within the framework of deep water Li, Hui \& Donelan \cite{lihd} studied 
the side-band instability of a Stokes wave train in uniform velocity shear. 
The coefficient of the nonlinear term of the NLS equation they derived was 
erroneous as noted by Baumstein \cite{baumstein}. The latter author investigated 
the effect of piecewise-linear velocity profiles in water of infinite depth on 
side-band instability of a finite-amplitude gravity wave. The coefficients of 
the NLS equation he derived were computed numerically because he did not give 
their expression as a function of the vorticity and depth of the shear layer,
explicitly. Instead, he calculated these coefficients for specific values of 
the vorticity and depth of shear layer.
Choi \cite{choi} considered the 
Benjamin-Feir instability of a modulated wave train in both positive and negative shear currents within the framework 
of the fully nonlinear water wave equations. For a fixed wave steepness, he compared his results with the irrotational 
case and found that the envelope of the modulated wave train grows faster in a positive shear current and slower 
in a negative shear current. Using the fully nonlinear equations, Okamura \& Oikawa \cite{okaoik} investigated 
numerically some instability characteristics of two-dimensional finite amplitude surface waves on a linear 
shearing flow to three-dimensional infinitesimal rotational disturbances.

The present study deals with the modulational instability of one dimensional, periodic water waves propagating 
on a vertically uniform shear current. We assume that the shear current has been produced by external effects 
and that the fluid is inviscid. In section~\ref{derivation} a NLS equation (vor-NLS equation) for surface waves propagating 
on finite depth in the presence of non zero constant vorticity is derived by using the method of multiple scales. 
In subsection~\ref{infinite} it is shown that the heuristic method to derive a 
NLS equation from a nonlinear
dispersion relation is not valid when vorticity is present.
This is a consequence of the coupling between the mean flow due to the modulation and the vorticity.
Section~\ref{stability} is devoted to a detailed stability analysis of a weakly nonlinear wave train as a function 
of the  parameter~$kh$  where $k$ is the carrier wavenumber and $h$ the depth and of vorticity magnitude. 
Consequences on  the Benjamin-Feir index are considered, too and a conclusion is given in section~\ref{conclusion}.

\section{Derivation of the vor-NLS equation}\label{derivation}
\begin{figure}
\begin{minipage}[b]{.46\linewidth}
\includegraphics[width=\linewidth]{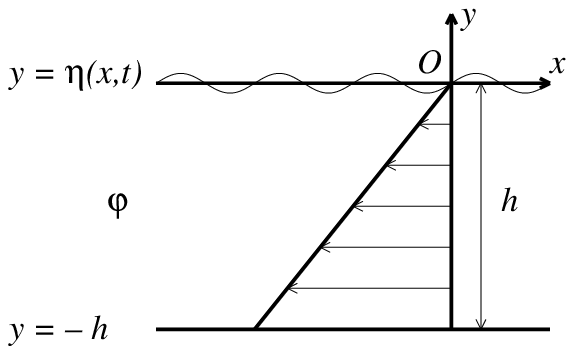}
\caption{Shear flow with $\Omega>0$ \hfill\break\hfill(waves propagating downstream)\label{fig1}}
\end{minipage} \hfill
\begin{minipage}[b]{.46\linewidth}
\includegraphics[width=\linewidth]{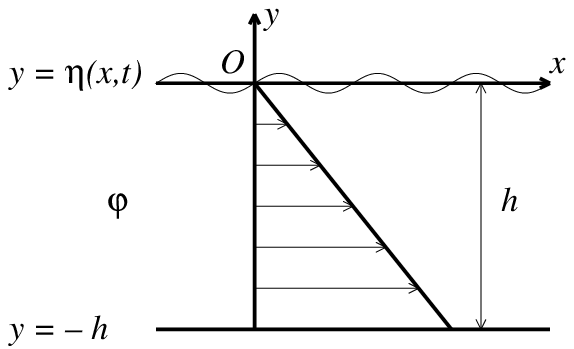}
\caption{Shear flow with $\Omega<0$ \hfill\break\hfill(waves propagating upstream)\label{fig2}}
\end{minipage}
\end{figure}

The undisturbed flow is a weakly nonlinear Stokes wave train propagating steadily 
on a shear current that varies linearly in the vertical direction $y$. 
The wave train moves along the $x$-axis. The $y$-axis is oriented upward, and gravity downward.
Naturally, $\bf i$ and $\bf j$ are unit vectors along $Ox$ and $Oy$.
When computing vector products, we shall also use \hbox{$\bf k=\bf i\wedge \bf j$}.
The depth $h$ is constant and the bed is located at \hbox{$y=-h$}.
Let $\Omega$ be the magnitude of the shear. There is a potential $\varphi(x,y,t)$ such that the 
velocity writes
\begin{equation} 
{\bf V}=\Omega y {\bf i} + \nabla \varphi(x,y,t)\label{champvit}
\end{equation} 
since for a two dimensional flow of an inviscid and incompressible fluid with external forces deriving from a 
potential the Kelvin theorem states that the vorticity is conserved. 
The variable $t$ is the time and $-\Omega$ is the vorticity in all the fluid that can be negative 
or positive as illustrated in figures~\ref{fig1}
and~\ref{fig2}, respectively. Note that the reference frame is in uniform translation with regard to that 
of the laboratory. Hence, the velocity of the undisturbed flow vanishes at the surface.
\subsection{Governing equations}
As the perturbation is assumed potential, the incompressibility condition \hbox{$\nabla . \vec V=0$}
implies that the velocity potential satisfies the Laplace's equation
\begin{equation}
\Delta \varphi =0\qquad -h < y < \eta(x,t)
\end{equation} 
The fluid is inviscid and so the Euler's equation writes~:
\begin{equation} 
\nabla(\varphi_t +\frac{1}{2}{\bf V}^2+\frac{P}{\rho}+gy)={\bf V}\wedge {\bf\omega} \label{PFD}
\end{equation} 
where $\omega$ is the vorticity vector~: \hbox{$\omega=-\Omega {\bf k}$}.
\newline
We introduce the stream function~$\psi$ associated to the velocity potential through the Cauchy-Riemann
relations~:
\begin{equation} 
\psi_y=\varphi_x,\qquad \psi_x=-\varphi_y \label{CauchyRiemann}
\end{equation} 
We notice that
\begin{equation} 
{\bf V}\wedge \omega=\nabla(\frac{1}{2}\Omega^2y^2+\Omega\psi)
\end{equation} 
so, we can rewrite equation (\ref{PFD}) as follows
\begin{equation}
\nabla(\varphi_t +\frac{1}{2}\varphi_x^2+\frac{1}{2}\varphi_y^2+\Omega y\phi_x+\frac{P}{\rho}+gy-\Omega\psi)=0.
\end{equation}
This equation may be integrated once~:
\begin{equation} 
\varphi_t +\frac{1}{2}\varphi_x^2+\frac{1}{2}\varphi_y^2+\Omega y\phi_x+\frac{P}{\rho}+gy-\Omega\psi=f(t)\label{bernbrut}
\end{equation} 
We write this equation at the free surface of the fluid. The notation~\hbox{$\Phi$} means that $\varphi$ 
is calculated on the free surface. The same convention is used for the derivatives of $\varphi$ or~$\psi$.
This convention will be used in the whole paper.

The pressure on the free surface is the atmospheric pressure that can be considered as a constant, 
and incorporated in the RHS of equation~(\ref{bernbrut}).

It is possible to add to the velocity potential function a primitive of the right hand side~$f(t)$
of this equation, so that this term vanishes.
The equation becomes
\begin{equation}
\Phi_t +\frac{1}{2}\Phi_x^2+\frac{1}{2}\Phi_y^2+\Omega \eta\Phi_x+g\eta-\Omega\Psi=0.
\end{equation}
\newline
The kinematic condition is written as follows
\begin{equation}
\Phi_y=\eta_x (\Phi_x+\Omega \eta)+\eta_t
\end{equation} 
\newline
The governing equations are then
\begin{eqnarray}
\nabla^2 \varphi=0, \quad  \quad -h<y<\eta(x,t) \label{Laplace} \\
\varphi_y =0, \hspace{2cm}\quad     \quad y=-h \label{fond}\\
\eta_t +(\Phi_x +\Omega\eta)\eta_x -\Phi_y =0  \label{cin}\\
\Phi_t +\frac{1}{2}\Phi_x^2+\frac{1}{2}\Phi_y^2+\Omega \eta\Phi_x+g\eta-\Omega\Psi=0
 \label{dyn}
\end{eqnarray}
\newline
To reduce the number of dependent variables, we derive the dynamic condition with respect to~$x$ 
and we use the Cauchy-Riemann conditions to eliminate the stream function. The result is
\begin{equation} 
\begin{split} 
\Phi_{tx}+\Phi_{ty}\eta_x+\Phi_x (\Phi_{xx}+\Phi_{xy}\eta_x)+\Phi_y (\Phi_{xy}+\Phi_{yy}\eta_x)
\\+\Omega \eta_x\Phi_x+\Omega \eta (\Phi_{xx}+\Phi_{xy}\eta_x)+g\eta_x+\Omega (\Phi_y-\Phi_x\eta_x)=0
\end{split} 
\end{equation} 
Equations (\ref{Laplace})-(\ref{dyn}) are invariant under the following transformations~: $\varphi \rightarrow 
-\varphi$, $t \rightarrow -t$, $\Omega \rightarrow -\Omega$ and $\Psi \rightarrow -\Psi$. 
Hence, there is no loss of generality if the study is restricted to waves with positive phase speeds 
so long as both positive and negative values of $\Omega$ are considered.

\subsection{The multiple scale analysis}
We seek an asymptotic solution in the following form
\begin{equation} 
\varphi=\sum_{n=-\infty}^{+\infty} \varphi_n \exp[in(kx-\omega t)],\qquad 
\eta=\sum_{n=-\infty}^{+\infty} \eta_n \exp[in(kx-\omega t)]
\end{equation} 
where $k$ is the wavenumber of the carrier and $\omega$ its frequency.
\newline
We assume $\varphi_{-n}=\varphi_n^*$ and $\eta_{-n}=\eta_n^*$ so that $\varphi$ and $\eta$ are real functions.
\newline
Then $\varphi_n$ and $\eta_n$ are written in perturbation series 
\begin{equation} 
\varphi_n=\sum_{j=n}^{+\infty}\epsilon^j \varphi_{nj},
\qquad \eta_n=\sum_{j=n}^{+\infty}\epsilon^j \eta_{nj}
\end{equation} 
where the small parameter $\epsilon$ is the wave steepness.
\newline 
We assume \hbox{$\varphi_{00}=0$} and \hbox{$\eta_{00}=0$}.\vspace{0.2cm}
\newline
Following Davey \& Stewartson \cite{davste}, we consider a solution that is modulated on the slow time scale 
$\tau=\epsilon^2 t$ and slow space scale $\xi=\epsilon(x-c_g t)$, where $c_g$ is the group velocity of the 
carrier wave.\vspace{0.3cm}
\newline
The new system of governing equations is 

\begin{align} 
&\epsilon^2\varphi_{\xi\xi}+\varphi_{yy}=0, \qquad -h\le y\le \eta(\xi,\tau)
\\& \varphi_y=0, \qquad y=-h
\\& \epsilon^2\eta_\tau-\epsilon c_g\eta_\xi +\epsilon^2\Phi_\xi\eta_\xi+\epsilon\Omega\eta\eta_\xi-\Phi_y =0
\\
\begin{split} 
\epsilon^3\Phi_{\xi\tau}&-\epsilon^2c_g\Phi_{\xi\xi}
+\epsilon^3\Phi_{y\tau}\eta_\xi-\epsilon^2 c_g\Phi_{\xi y}\eta_\xi
+\epsilon^3\Phi_\xi\Phi_{\xi\xi}
\\&
+\epsilon^3\Phi_\xi\Phi_{\xi y}\eta_\xi
+\epsilon\Phi_y\Phi_{\xi y}
+\epsilon\Phi_y\Phi_{yy}\eta_\xi
+\epsilon^2\Omega \eta_\xi\Phi_\xi
+\epsilon^2\Omega \eta\Phi_{\xi\xi}
\\&
+\epsilon^2\Omega \eta\Phi_{\xi y}\eta_\xi
+\epsilon g\eta_\xi
+\Omega\Phi_y
-\epsilon^2\Omega\Phi_\xi\eta_\xi=0
\end{split} 
\end{align}
Substituting the expansions for the potential $\varphi$ into the Laplace equation and using the
method of multiple scales we obtain
\begin{align}  
\varphi_{01yy}&=0
\\ -k^2 \varphi_{11}+\varphi_{11yy}&=0
\\ \varphi_{02yy}&=0
\\ -k^2 \varphi_{12}+2 ik \varphi_{11\xi}+\varphi_{12yy}&=0
\\ -4 k^2 \varphi_{22}+\varphi_{22yy}&=0
\\ \varphi_{01\xi\xi}+\varphi_{03yy}&=0
\\ -k^2 \varphi_{13}+2 ik \varphi_{12\xi}+\varphi_{11\xi\xi}+\varphi_{13yy} &=0
\\ -4 k^2 \varphi_{23}+4 ik \varphi_{22\xi}+\varphi_{23yy}&=0
\\ -9 k^2 \varphi_{33}+\varphi_{33yy}&=0
\end{align} 
Solving these equations and considering the bottom conditions we obtain~:
\begin{align} 
\varphi_{01y}&=0\\
\varphi_{02y}&=0\\
\varphi_{11}&=A\frac{\cosh[k(y+h)]}{\cosh(kh)}\\
\varphi_{12}&=D\frac{\cosh[k(y+h)]}{\cosh(kh)}-iA_\xi\frac{(y+h)\sinh[k(y+h)]-h\sigma\cosh[k(y+h)]}{\cosh(kh)}\\
\varphi_{22}&=F\frac{\cosh[2k(y+h)]}{\cosh(2kh)}\\
\varphi_{03y}&=-(y+h)\phi_{01\xi\xi}\\
\begin{split} 
\varphi_{13}&=(h\sigma A_{\xi\xi}-iD_\xi)\frac{(y+h)\sinh[k(y+h)]}{\cosh(kh)}
\\&+(B+\frac{h^2}2(1-2\tanh^2(kh))A_{\xi\xi}+ih\sigma D_\xi-A_{\xi\xi}\frac{(y+h)^2}2)
\frac{\cosh[k(y+h)]}{\cosh(kh)}
\end{split} 
\end{align} 
The next tedious step is to use the relations obtained from the kinematic and dynamic conditions.
Let us set $\oOmega=\frac{\Omega}{\omega}$. Herein, it is important to emphasize that this parameter 
does not correspond to a dimensionless vorticity because the frequency $\omega$ depends on $\Omega$.
Furthermore, we set $X=\sigma\oOmega$, where \hbox{$\sigma=\tanh(kh)$}, 
because this term will occur many times in the following polynomial expressions.
\begin{itemize}
\item Terms in $\epsilon E^0$~: They give no supplementary information
\item Terms in $\epsilon E^1$~: They give a linear dispersion relation
\begin{equation} 
kc_p^2+\sigma(c_p\Omega-g)=0\label{reldisplin}
\end{equation} 
From this linear dispersion relation it is easy to demonstrate that we have always $X>-1$ or 
$\oOmega>-1/\sigma$.
\newline
We also get a relation between the fundamental modes of the velocity potential at the surface 
and free surface elevation~:
\begin{equation} 
\eta_{11}=i\frac{\sigma}{c_p}A=\frac{i\omega}{g}(1+X)A\label{propo}
\end{equation} 
where $c_p=\omega/k$.
\newline
From the linear dispersion relation the group velocity is~:
\begin{equation} 
c_g=\frac{c_p}{\sigma}\times\frac{(1-\sigma^2)kh+\sigma(1+X)}{2+X}
\end{equation} 
We recall that \hbox{$X>-1$}, so there is no singularity in this expression.
\item Terms in $\epsilon^2 E^0$~: They only give, after simplifications~:
\begin{equation}
\eta_{01\xi}=0
\end{equation}
\item Terms in $\epsilon^2 E^1$~: They give a system of two equations with two indeterminate
coefficients $\eta_{01}$ and $\eta_{12}$ that can be found after some calculations~:
\begin{equation}
\eta_{01}=0
\end{equation} 
and 
\begin{equation}
\eta_{12}=\frac{1}{g}[c_g+h(1-\sigma^2)\Omega]A_\xi+\frac{i\omega}{g}(1+X)D
\end{equation} 
\item Terms in $\epsilon^2 E^2$~: They give a system of two linear equations with $F$ and $\eta_{22}$ as unknowns~:
\begin{equation} 
F=i\omega(1+\sigma^2)\frac{3(1-\sigma^2)+3X+X^2}{4\sigma^2 c_p^2}A^2
\end{equation} 
and
\begin{equation}
\eta_{22}=-\frac{k}{2c_p^2\sigma}[3-\sigma^2+(3+\sigma^2)X+X^2] A^2
\end{equation} 
\item Terms in $\epsilon^3 E^0$~: 
The first-order mean flow can be obtained from the following expression
\begin{equation} 
[c_g(c_g+\Omega h)-gh] \varphi_{01\xi}=\ec{\frac{g\sigma\omega}{c_p^2}(2+X)+k^2 c_g(1-\sigma^2)}\abs A^2
\end{equation} 
and 
\begin{equation} 
g\eta_{02}=(c_g+\Omega h)\phi_{01\xi}-k^2(1-\sigma^2) \abs A^2
\end{equation} 
\item Terms in $\epsilon^3 E^1$~: We derive two equations from which it is possible, 
after tedious computations, to eliminate $\eta_{13}$. 
The coefficients $B$ and $D_\xi$ vanish owing to the linear dispersion relation. 
The remaining terms are~: A time derivative, a dispersive term, a nonlinear term 
and a term involving the mean flow that we can substitute by its expression
taken from the other equation. 
Finally a nonlinear Schr\"{o}dinger equation with vorticity is derived (the vor-NLS equation)

\begin{equation}
iA_\tau+L A_{\xi\xi}=P \mid A \mid^2 A\label{vnlspot}
\end{equation} 
where
\begin{align} 
L&=\frac{\omega}{k^2\sigma(2+X)}{\mu(1-\sigma^2)[1-\mu\sigma+(1-\rho)X]-\sigma\rho^2}
\\
P&=\frac{k^4c_p}{2(2+X)g\sigma^3} (U+VW)=\frac{k^4(U+VW)}{2(1+X)(2+X)\omega\sigma^2} 
\\
\begin{split} 
U&= 9-12\sigma^2+13\sigma^4-2 \sigma^6
+(27-18\sigma^2+15\sigma^4)X
\\&
+(33-3\sigma^2+4\sigma^4)X^2
+(21+5\sigma^2)X^3
+(7+2\sigma^2)X^4
+X^5
\end{split} \label{defU}
\\
V&=(1+X)^2(1+\rho+\mu\oOmega)+1+X-\rho\sigma^2-\mu\sigma X \label{defV} \\
W&=2\sigma^3 \frac{(1+X)(2+X)+\rho(1-\sigma^2)}{\sigma \rho(\rho+\mu\oOmega)-\mu(1+X)} \label{defW}
\end{align} 
\end{itemize}
with
\begin{align}
\mu&=kh\\
\sigma&=\tanh(\mu)\\
\rho&=\frac{c_g}{c_p}\quad (\text{not to be confused with the density})
\end{align} 
The relation~(\ref{propo}) permits to replace the velocity potential~$A$
by the elevation~$a$.
\begin{equation} 
ia_\tau+ L a_{\xi\xi}=M \mid a \mid^2 a \label{vnlseta}
\end{equation} 
where $a$ is the envelope of the surface elevation and
\begin{equation} 
M=\frac{\omega k^2(U+VW)}{8(1+X)(2+X)\sigma^4}
\end{equation} 

\subsection{The case of infinite depth}\label{infinite}
Let us discuss what happens when depth goes to infinity in order 
to compare our results to those of 
Li {\em et al}.~\cite{lihd} or Baumstein~\cite{baumstein}.
Moreover, we shall show the importance of the coupling
between the mean flow and vorticity at third order.
At this order before deriving equation~(\ref{vnlspot}) the following coupled equations 
empasize the coupling between the mean flow~\hbox{$\phi_{01\xi}$} and vorticity~$\oOmega$.
\begin{equation}
\begin{split}
&\frac{k ^3c_p^2}{g\sigma}
\Big[(1+\sigma\oOmega)^2\ep{c_p+c_g+k hc_p\oOmega}+c_p(1+\sigma\oOmega)
\\&\phantom{\frac{k ^3c_p^2}{g\sigma}\Big[}
-(c_g+c_pk h\oOmega)\sigma^2\big]
\phi_{01\xi}A
-i\omega  (2+\sigma\oOmega) A_\tau
\\&+\ea{c_g^2-gh+gh\sigma\ec{\sigma+k h(1-\sigma^2)}+c_pk h(1-\sigma^2)\ep{c_g-c_pk h\sigma}\oOmega}A_{\xi\xi}
\\&=-\frac{k ^5c_p^2}{2g\sigma^3}
\Big[9-12\sigma^2+13\sigma^4-2 \sigma^6
+(27-18\sigma^2+15\sigma^4)\sigma\oOmega
\\&\phantom{=-\frac{k ^5c_p^2}{2g\sigma^3}\Big[}
+(33-3\sigma^2+4\sigma^4)\sigma^2\oOmega^2
+(21+5\sigma^2)\sigma^3\oOmega^3
\\&\phantom{=-\frac{k ^5c_p^2}{2g\sigma^3}\Big[}
+(7+2\sigma^2)\sigma^4\oOmega^4
+\sigma^5\oOmega^5
\Big]\abs A^2 A\label{bbvnls}
\end{split}
\end{equation}
with
\begin{equation}
\phi_{01\xi}=\frac{gk \sigma\ep{2+\sigma\oOmega}+k ^2c_pc_g(1-\sigma^2)}{c_p\ec{c_g\ep{c_g+\Omega h}-gh}}\abs A^2\label{meanflow}
\end{equation}

The mean flow~\hbox{$\phi_{01\xi}$} verifies
\begin{equation}
\lim_{h\to+\infty}h\phi_{01\xi}=\frac{gk \ep{2+\oOmega}}{c_p\ep{\Omega c_g-g}}\abs A^2
\end{equation}

The coefficient of the mean flow, induced by the modulation of the envelope,
in equation~(\ref{bbvnls}) is of order~\hbox{$O(h)$}
so that the product has a finite limit when \hbox{$h\to\infty$}. More precisely,
the coefficient of~\hbox{$h\phi_{01\xi}A$} in equation~(\ref{bbvnls}) goes to
\begin{equation}
\frac{k ^4c_p^3\oOmega^2}{g}\ep{2+\oOmega}
\end{equation}
when \hbox{$h\to\infty$}.

The remaining terms in equation~(\ref{bbvnls}) 
have also finite limits when \hbox{$h\to\infty$}.
Finally, we get the following NLS equation valid for infinite depth 
and constant vorticity~:
\begin{equation}
iA_\tau
-\frac{\omega(1+\oOmega)^2}{k ^2(2+\oOmega)^3} A_{\xi\xi}
=
-\frac{\omega k ^2}{2c_p^2}\frac{\oOmega^2\ep{2+\oOmega}^2}{1+\oOmega} \abs A^2 A
+\frac{\omega k^2}{2c_p^2}
\ep{4+6\oOmega+6\oOmega^2+\oOmega^3}\abs A^2 A
\end{equation}
This equation deals with~$A$ which is the value of the velocity potential
for~\hbox{$y=0$}. 
Using equation~(\ref{propo}), the NLS equation for the enveloppe of the wavetrain is~:
\begin{equation}
ia_\tau
-\frac{\omega(1+\oOmega)^2}{k ^2(2+\oOmega)^3} a_{\xi\xi}
=
-\frac{\omega k ^2}{8}\frac{\oOmega^2\ep{2+\oOmega}^2}{1+\oOmega} \abs a^2 a
+\frac{\omega k^2}{8}
\ep{4+6\oOmega+6\oOmega^2+\oOmega^3}\abs a^2 a\label{nlssep}
\end{equation}
We have left deliberately two nonlinear terms.
The first term of the RHS comes from
the coupling between the mean flow and the vorticity
while the second can be obtained heuristically from
the nonlinear dispersion relation.
When $\oOmega$ vanishes, this coupling disappears and the heuristic method
can be applied. Note that
Baumstein\cite{baumstein} and Li {\em et al.}\cite{lihd} missed this coupling.

If we use the heuristic method to obtain the NLS equation from the 
nonlinear dispersion relation that was found by Simmen \& Saffman \cite{simsaf},
we should obtain only the second term of the RHS of~(\ref{nlssep}).
So, we have shown that the heuristic method is not valid
in presence of vorticity, even in infinite depth.

Equation~(\ref{nlssep}) is rewritten as follows~:
\begin{equation}
ia_\tau -\frac{\omega(1+\oOmega)^2}{k ^2(2+\oOmega)^3} a_{\xi\xi} =
\frac{\omega k^2}{8(1+\oOmega)}
\ep{4+10\oOmega+8\oOmega^2+3\oOmega^3}\abs a^2 a
\label{vornlsinf}
\end{equation}

The dispersive and nonlinear coefficients of equation (\ref{vornlsinf}) 
present two poles $\oOmega=-2$ 
and $\oOmega=-1$ and two zeros $\oOmega=-1$ and $\oOmega=-2/3$ respectively. 
Nevertheless, we recall that $\oOmega>-1$ and consequently in infinite depth 
$\oOmega$ will never be equal to $-1$ or $-2$. 
\newline
For $\oOmega=-2/3$, the nonlinear coefficient vanishes and the vor-NLS equation is reduced to a linear dispersive 
equation (the Schr\"{o}dinger equation)
	\begin{equation}
	ia_{\tau}-\frac{3}{64}a_{\xi \xi}=0
	\end{equation}
\section{Stability analysis and results}\label{stability}
\begin{figure}
  \centering
        \includegraphics[width=0.8\textwidth]{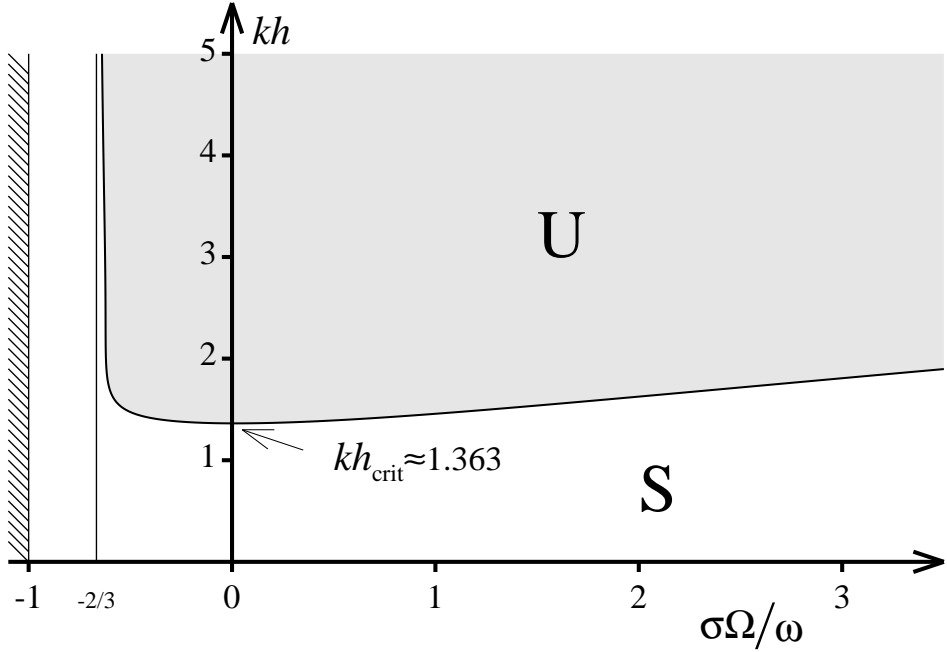}
       \caption{Stability diagram in the $(\oOmega,kh)$-plane. {\bf{S}}~: stable, {\bf{U}}~: unstable.} 
   \label{graphstable}
\end{figure}

\begin{figure}
\begin{minipage}[b]{.46\linewidth}
\includegraphics[width=\linewidth]{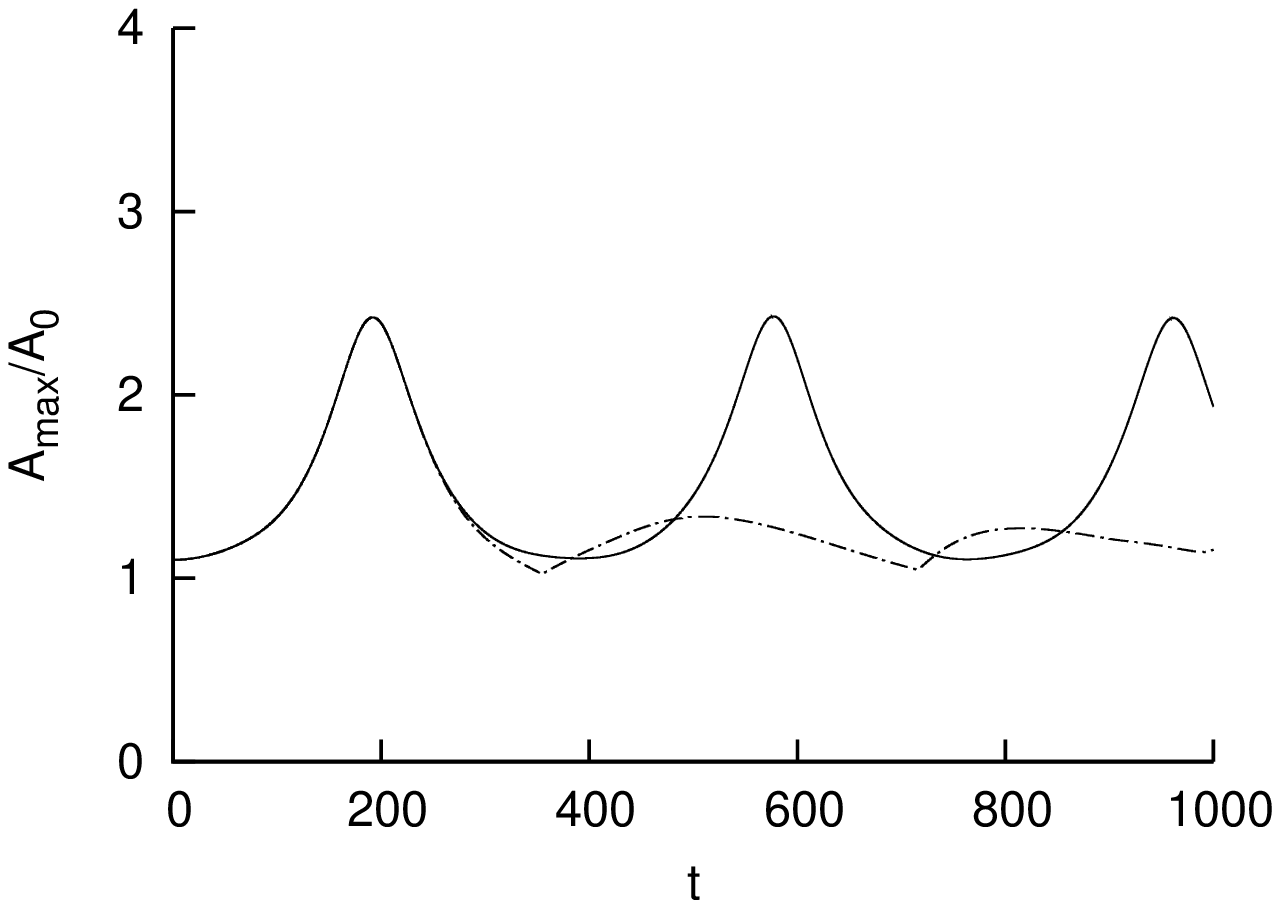}
\caption{Temporal evolution of the normalized maximum amplitude of the envelope 
in the case of a simple recurrence for $kh=\infty$~: 
\hbox{$\oOmega=0$} (solid line), $\oOmega=-0.83$ (dash-dotted line)}
\label{restab1}
\end{minipage} \hfill
\begin{minipage}[b]{.46\linewidth}
\includegraphics[width=\linewidth]{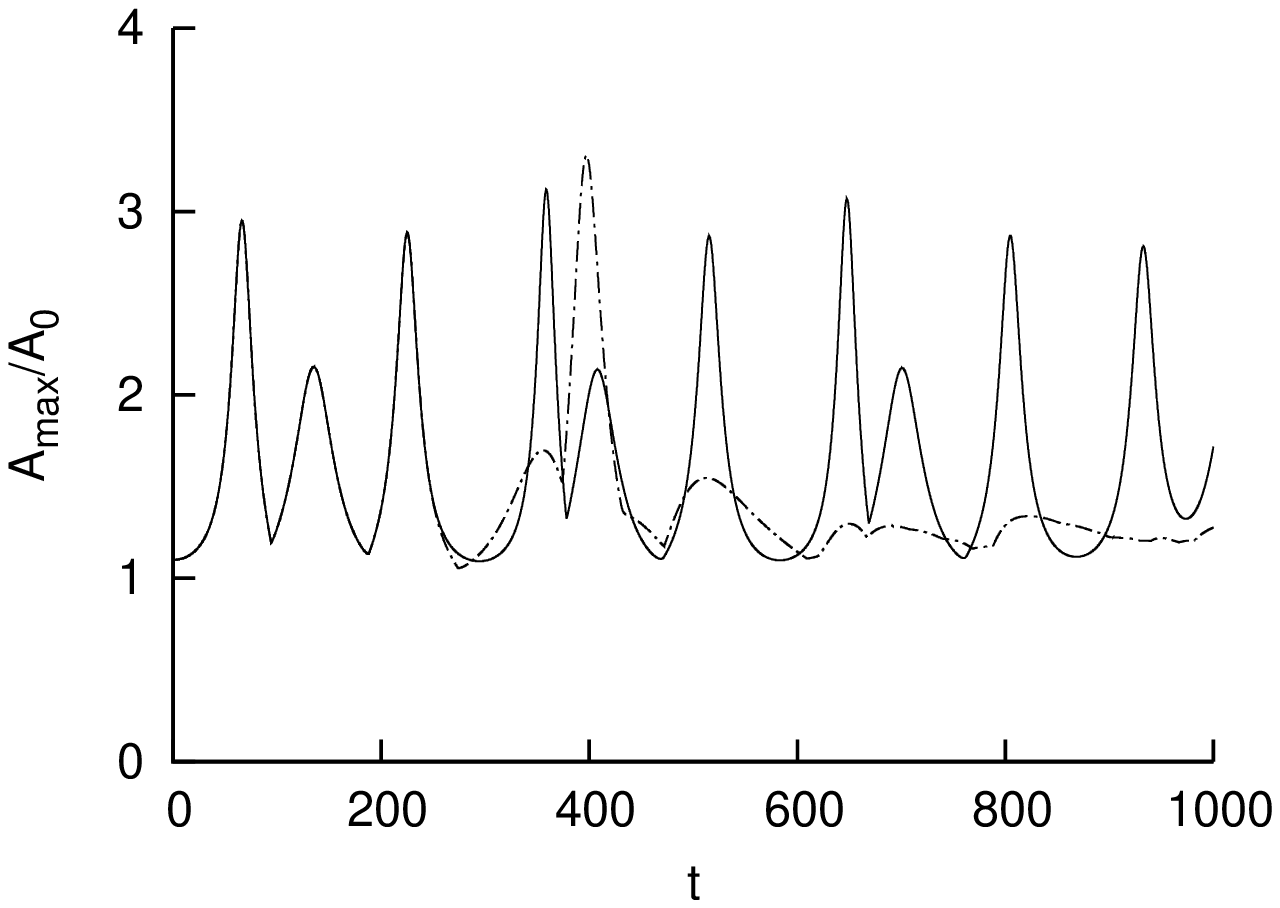}
\caption{Temporal evolution of the normalized maximum amplitude of the envelope 
in the case of a double recurrence for $kh=\infty$~: 
\hbox{$\oOmega=0$} (solid line), $\oOmega=-0.83$ (dash-dotted line)}
\label{restab2}
\end{minipage}
\end{figure}

The equation (\ref{vnlseta}) admits the following Stokes's wave solution
\begin{equation}
a=a_0 \exp(-iMa_0^2\tau)
\end{equation}
We consider the following infinitesimal perturbation of this solution
\begin{equation}
a=a_0(1+\delta_a) \exp[i(\delta_\omega-Ma_0^2\tau)]
\end{equation}
Substituting this expression in equation (\ref{vnlseta}) and linearizing about the Stokes' wave solution, we obtain
\begin{equation}
i\frac{\partial\delta_a}{\partial \tau}-\frac{\partial\delta_\omega}{\partial \tau}+\delta_a Ma_0^2
+L\frac{\partial^2\delta_a}{\partial\xi^2}+iL\frac{\partial^2\delta_\omega}{\partial\xi^2}-3Ma_0^2 \delta_a=0
\end{equation}
Separating the real and imaginary parts, the previous equation transforms into the following system
\begin{equation}
\left\{
\begin{aligned}
&\frac{\partial\delta_a}{\partial \tau}
+L\frac{\partial^2\delta_\omega}{\partial\xi^2}
=0
\\
&
L\frac{\partial^2\delta_a}{\partial\xi^2}
-2Ma_0^2 \delta_a
-\frac{\partial\delta_\omega}{\partial \tau}
=0
\end{aligned} 
\right.
\label{systlin}
\end{equation} 
This is a system of linear differential equations with constant coefficients that admits
the following solution 
\begin{equation}
\left\{
\begin{aligned}
\delta_a&=\Delta_a\exp[i(l\xi-\lambda\tau)]
\\ \delta_\omega&=\Delta_\omega\exp[i(l\xi-\lambda\tau)]
\end{aligned} 
\right.
\end{equation} 
Substituting this solution in the system of equations (\ref{systlin}) gives
\begin{equation}
\left\{
\begin{aligned}
i\lambda \Delta_a & + & l^2 L \Delta_\omega &=& 0\\
(2Ma_0^2+l^2 L)\Delta_a & - & i\lambda \Delta_\omega &=& 0
\end{aligned} 
\right.
\end{equation} 
The necessary and sufficient condition of non trivial solutions is~:
\begin{equation}
\lambda^2=\ell^2 L (2Ma_0^2+\ell^2 L)
\end{equation} 
Discussion~: When $L (2Ma_0^2+\ell^2 L)\ge 0$ there are two real solutions,
the perturbation is bounded and the Stokes' wave solution is stable while
when $L (2Ma_0^2+l^2 L)<0$ the perturbation is unbounded and the solution is unstable. 
Note that the latter condition implies that $LM<0$.
\newline
We set $L=L_1\frac{\omega }{k^2}$ and $M=M_1\omega k^2$
so that $L_1$ and $M_1$ are dimensionless functions of $kh$ and~$\oOmega$ only.
The growth rate of instability is then 
\begin{equation}
\gamma = \frac{l\omega }{k^2}\sqrt{-2M_1L_1k^4a_0^2-l^2L_1^2}\label{taux}
\end{equation} 
\newline
Its maximal value is obtained for $l=\sqrt{-\frac{M_1}{L_1}}a_0k^2$
and is~:
\begin{equation}
\gamma_{\hbox{\tiny max}}=M_1\omega (ka_0)^2
\label{growthmx}
\end{equation}
The instability domain is plotted in figure \ref{graphstable} as a function of the parameters 
$\oOmega$ and~\hbox{$kh$}. As soon as $-1/\sigma<\oOmega\leq -2/3$ the waves become stable to 
modulational perturbations. Noting that $\oOmega$ is an increasing function of~$\Omega$,
it is easy to show that $-1/\sigma<\oOmega \leq -2/3$ corresponds to $-\infty<\Omega \leq -2\sqrt{\frac{kg}{3}}$. 
Hence, there is a value~$\Omega_c=-2\sqrt{\frac{kg}{3}}$ of~$\Omega$ (depending on~$k$) for which \hbox{$\oOmega=-2/3$}. 
For a constant vorticity corresponding to \hbox{$\Omega \leq \Omega_c$} waves are linearly stable. 
In particular, Stokes' waves of wavenumber $k$ propagating on a linear shear current satisfying 
$\Omega \leq -2\sqrt{\frac{kg}{3}}$ are stable to modulational instability whatever the value of 
the depth may be. 
\newline
There is a critical value $kh_{\mathrm{crit}}$ of the parameter $kh$, 
as shown in figure~\ref{graphstable}, above which instability prevails. 
For \hbox{$\Omega=0$} (no vorticity) this threshold has the well known value~1.363. 
The critical value of this threshold is reached very near \hbox{$\Omega=0$}.
\newline
The linear stability of the Stokes wave solution is known to be controlled by the sign of the product $LM$ 
of the coefficients of the vor-NLS equation (\ref{vnlseta}). Let us consider this product when $kh \rightarrow
\infty$
\begin{equation}
LM=-\frac{\omega ^2}{8}\frac{(1+\oOmega)(2+3\oOmega)(2+2\oOmega+\oOmega^2)}{(2+\oOmega)^3}
\end{equation}
The condition $LM < 0$ corresponds to instability whereas $LM > 0$ corresponds to stability.
In the domain \hbox{$\oOmega>-1$}, this product admits one simple root \hbox{$\oOmega=-2/3$}.
For this value of $\oOmega$, $LM$ changes sign and as a result there is an exchange of stability. 
Hence, in infinite depth we can claim that there is no modulational instability when \hbox{$-1<\oOmega \leq -2/3$}.
\vspace{0.1cm}
\newline
In order to illustrate the restabilisation of the modulational instability 
we consider a modulated wave packet that propagates initially without current 
in infinite depth and meets progressively a current with $\oOmega=-0.83$ 
which corresponds to a stable regime. The results of the numerical simulations 
of the vor-NLS equation are shown in figures \ref{restab1} and \ref{restab2}. 
Temporal evolutions of the ratio $A_{\mathrm{max}}(t)/A_0$ are plotted without 
($\oOmega=0$) and with ($\oOmega=-0.83$) shear current  
where $A_{\mathrm{max}}(t)$ and $A_0$ are the maximum amplitudes of the modulated wave 
train at time $t$ and time $t=0$, respectively. In figure \ref{restab1} the vorticity is 
initially set equal to zero. At $t=200$ the value of  $\oOmega$ is increased progressively 
up to $-0.83$ (that belongs to $]-1,-2/3]$) at $t=600$ and remains equal to this value till 
the end of the numerical simulation. The carrier amplitude and carrier wavenumber are 
$kA_0=1/16$ and $k=8$ respectively. The perturbation amplitude is one tenth of the carrier 
amplitude and its wavenumber is $\Delta k=l=1$. Hence, the criterion for the occurrence 
of a simple recurrence is satisfied. For $\oOmega=0$, one can observe the Fermi-Pasta-Ulam 
recurrence phenomenon (FPU) which corresponds to a series of modulation-demodulation cycles. 
When the shear current is introduced the Benjamin-Feir instability is strongly reduced. 
In figure \ref{restab2}, the same numerical simulation is conducted, but the wave steepness 
of the carrier wave is now $kA_0=\frac{\sqrt{3}}{16}=0.1083$ and so the wavenumber $2l$ 
corresponds to an unstable perturbation. In figure \ref{restab2} is shown a double recurrence 
in the absence of shear current. The introduction of the vorticity modifies drastically this recurrence. 
When $\oOmega$ reaches the value $-0.83$, the modulational instability is removed. 
Note in the presence of vorticity the increase of the amplitude of the envelope of the wave packet 
near $t=400$. At this time $\oOmega$ does not yet belong to the stable interval $]-1, -2/3]$. 

\subsection{Growth rate of instability}
\begin{figure}
  \centering
\includegraphics[width=0.7\linewidth]{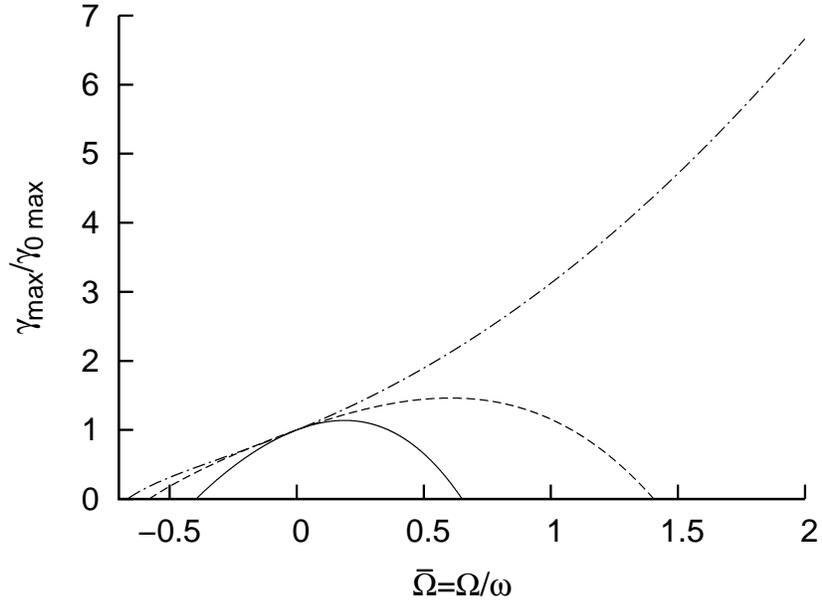}
\caption{Normalized maximum growth rate as a function of $\oOmega$ for $kh=1.40$ (solid line),
$kh=1.70$ (dashed line) and $kh=\infty$ (dash-dotted line). $\gamma_{0\hbox{\tiny{max}}}$ is the maximum 
growth rate in the absence of shear current}
\label{growth1}
\end{figure}

\begin{figure}
  \centering
\includegraphics[width=0.7\linewidth]{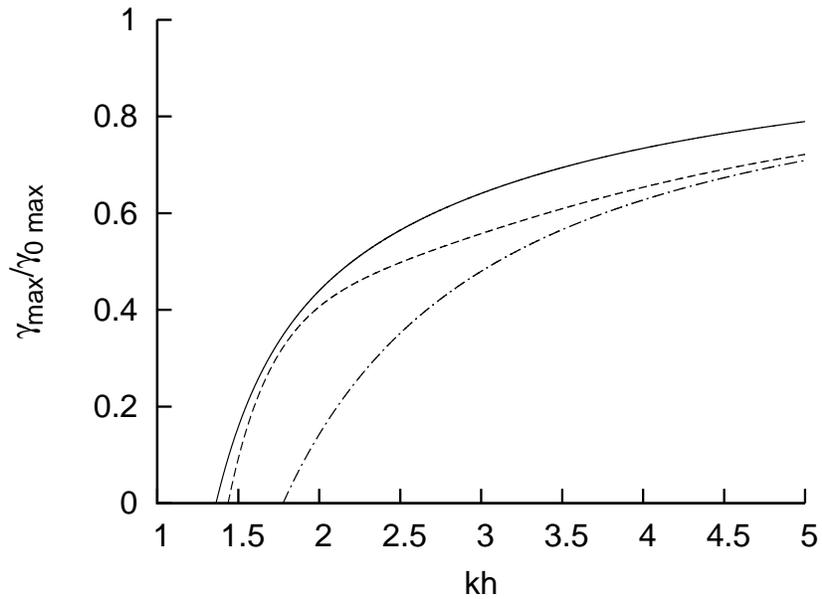}
\caption{Normalized maximum growth rate as a function of $kh$ for $\oOmega=0$ (solid line),
$\oOmega=-0.50$ (dashed line) and $\oOmega=3.0$ (dash-dotted line). 
$\gamma_{0 \hbox{\tiny{max}}}$~is the maximum growth rate when $kh=\infty$}
\label{growth3}
\end{figure}

The ratio of the maximum growth rate of instability given by equation (\ref{growthmx}) 
to its value in the absence of shear is plotted in figure \ref{growth1}  
as a function of $\oOmega$ for $\oOmega>-2/3$ and several values of $kh$. 
In infinite depth, the presence of vorticity increases or decreases the maximum growth rate 
of modulational instability, $\gamma_{\hbox{\tiny{max}}}$, when $\oOmega> 0$ or $-2/3 < \oOmega < 0$, respectively. 
In finite depth and $-2/3 < \oOmega < 0$, the effect of vorticity is to reduce the maximum rate 
of growth whereas for $\oOmega>0$ we observe an increase and then a decrease.
\newline
In figure \ref{growth3} is shown the behavior of the normalized maximum growth rate as a function of $kh$ 
for several values of $\oOmega$. Herein, the normalization is different from that used in figure \ref{growth1}. 
Figure \ref{growth3} correspond to values of $\oOmega$ larger than $-2/3$. 
For $\oOmega \geq -2/3$, the critical value $kh_{\mathrm{crit}}$ associated to restabilisation is very close to $1.363$ 
and corresponds to $\oOmega\approx 0$. 
In figure \ref{growth3} for $\oOmega > -2/3$ the maximum growth rate of instability increases 
with $kh$ greater than $1.363$.

\begin{figure}[ht]
\begin{minipage}[b]{.46\linewidth}
\includegraphics[width=\linewidth]{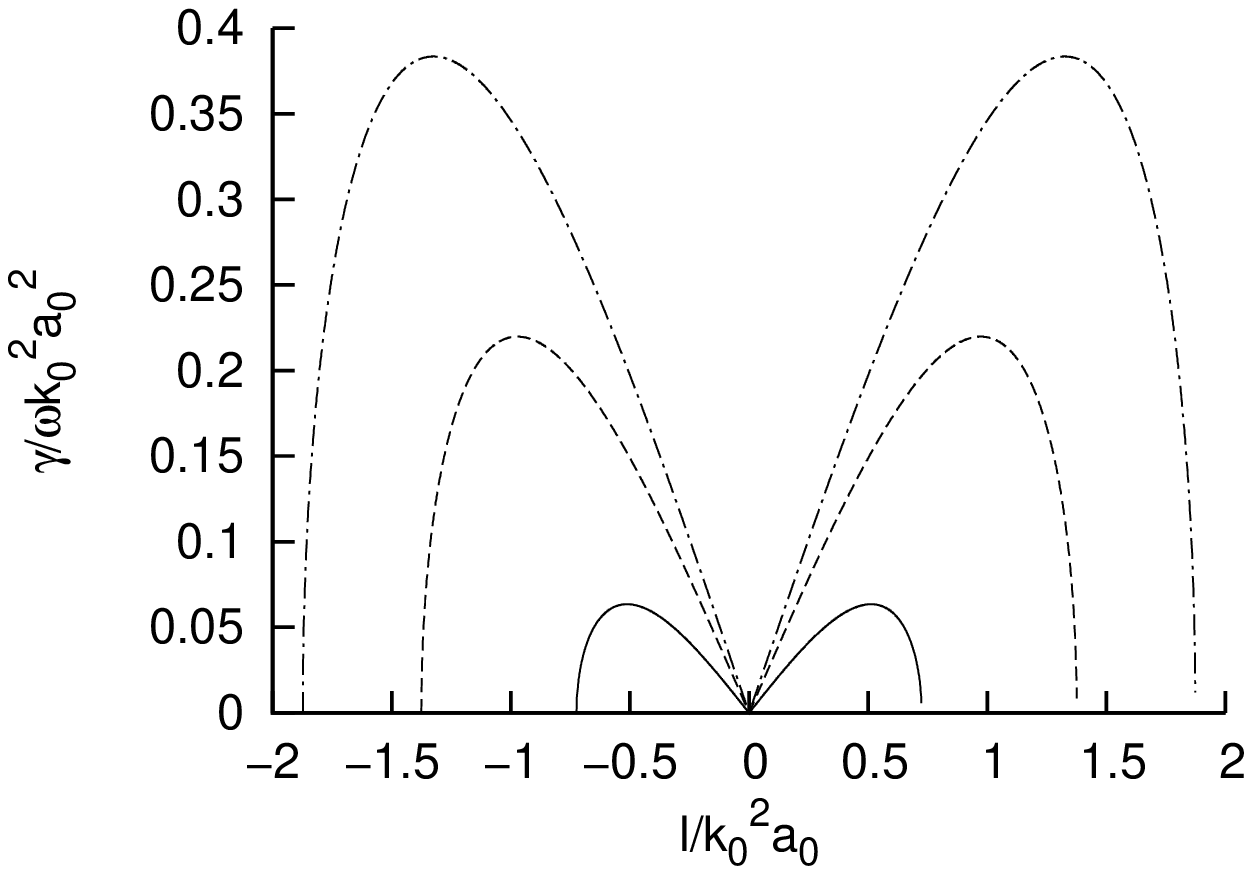}
\caption{Normalized growth rate as a function of the perturbation wavenumber $\ell$ for $kh=2.0$
and $\oOmega=-0.5$ (solid line), $\oOmega=0.0$ (dashed line),
$\oOmega=0.5$ (dot-dashed line)}
\label{growth5}
\end{minipage} \hfill
\begin{minipage}[b]{.46\linewidth}
\includegraphics[width=\linewidth]{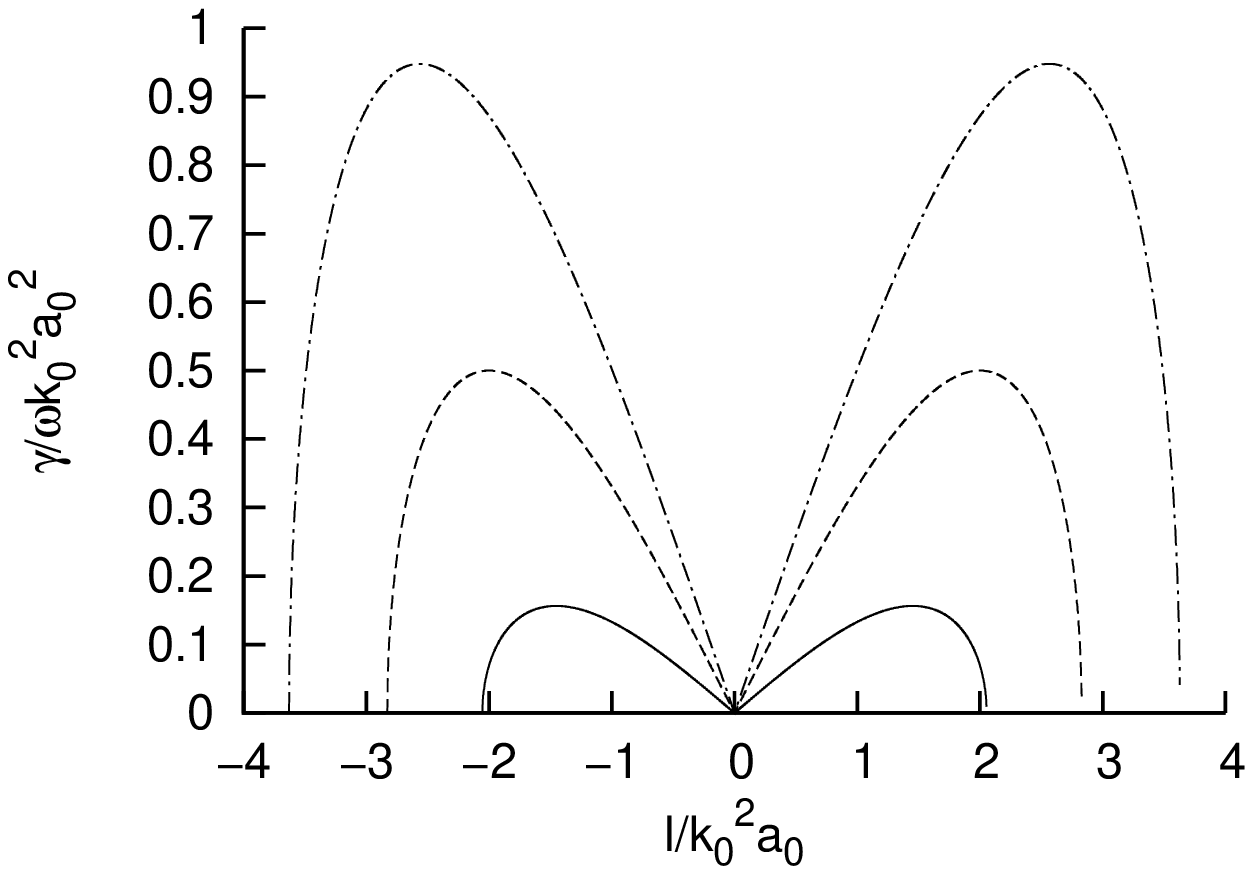}
\caption{Normalized growth rate as a function of the perturbation wavenumber $\ell$ for $kh=\infty$
and $\oOmega=-0.5$ (solid line), $\oOmega=0.0$ (dashed line),
$\oOmega=0.5$ (dot-dashed line)}
\label{growth7}
\end{minipage}
\end{figure}

In figure \ref{growth5} is plotted the normalized rate of growth of modulational instability 
as a function of the perturbation wavenumber $\ell$ for several values of $\oOmega$, 
within the framework of finite depth. Figure \ref{growth7} corresponds to the case of infinite depth. 

\subsection{Bandwidth instability}
\begin{figure}
  \centering
\includegraphics[width=0.7\linewidth]{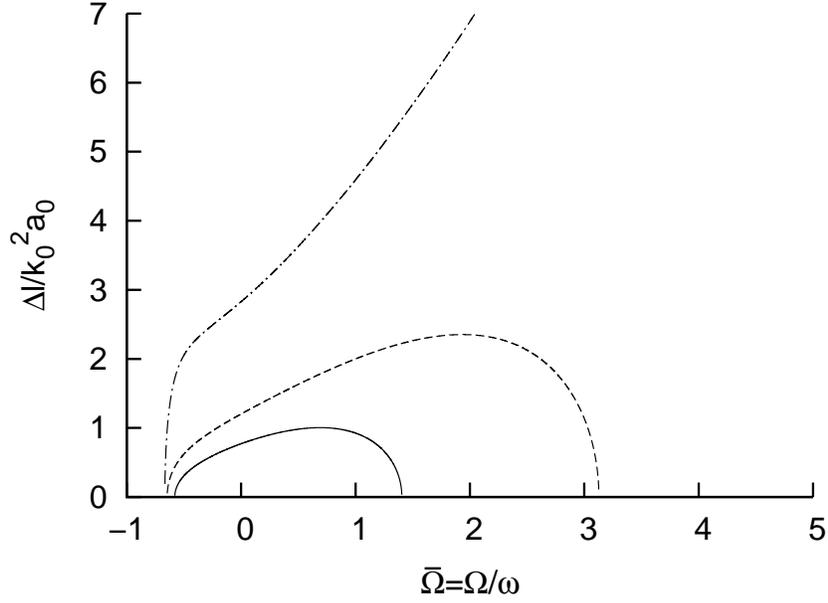}
\caption{Normalized instability bandwidth as a function of $\oOmega$ for $kh=1.5$ (solid line),
$kh=1.8$ (dashed line), $kh=\infty$ (dot-dashed line)}
\label{band}
\end{figure}

In figure \ref{band} is shown the ratio of the instability bandwidth $\Delta \ell$ 
to its value in the absence of shear current $\Delta \ell_0=\Delta \ell (\oOmega=0)$ 
as a function of $\oOmega$ for several values of $kh$.  
From equation (\ref{taux}), the instability bandwidth is $\sqrt{2\abs{\frac{M_1}{L_1}}}k^2a_0$. 
One can observe an increase of the band of instability followed by a decrease when $\oOmega$ increases, 
except when depth becomes infinite. 

In table 1 is presented a comparison of our results with those of Oikawa {\em et al.} (1987) 
in the case of two dimensional flows for two values of $kh$ and several values of the Froude number. 
Note that the Froude number, $F$, they used is exactly $\Omega$. 
This comparison shows a quite good agreement between Oikawa {\em et al.} and present results.

\begin{table}
\begin{center}
\begin{tabular}{|c|c|c|c|c|c|c|}
\hline
 & $F=0.0$ & $F=0.25$ & $F=0.5$ & $F=1.0$ & $F=1.5$ & $F=2.0$ \\ 
\hline
$kh=1.5$ & $1.6/1.54$ & $1.3/1.28$ & $1.0/1.00$ & $1.2/1.31$ & $6.0/5.96$ & $6.6/6.69$ \\
\hline
$kh=2.0$ & $2.8/2.75$ & $2.4/2.40$ & $2.0/1.97$ & $4.8/4.72$ & $-/-$      & $-/-$ \\
\hline
\end{tabular}
\caption{Comparison with results of Oikawa {\em et al.} (1987)~: 
$F$~is the Froude number.
The first value is estimated from their figures whereas the second one 
corresponds to our computations with the vor-NLS equation}
\end{center}
\end{table}

\subsection{Benjamin-Feir index in the presence of vorticity~: Application to rogue waves}
\begin{figure}
\begin{minipage}[b]{.46\linewidth}
\includegraphics[width=\linewidth]{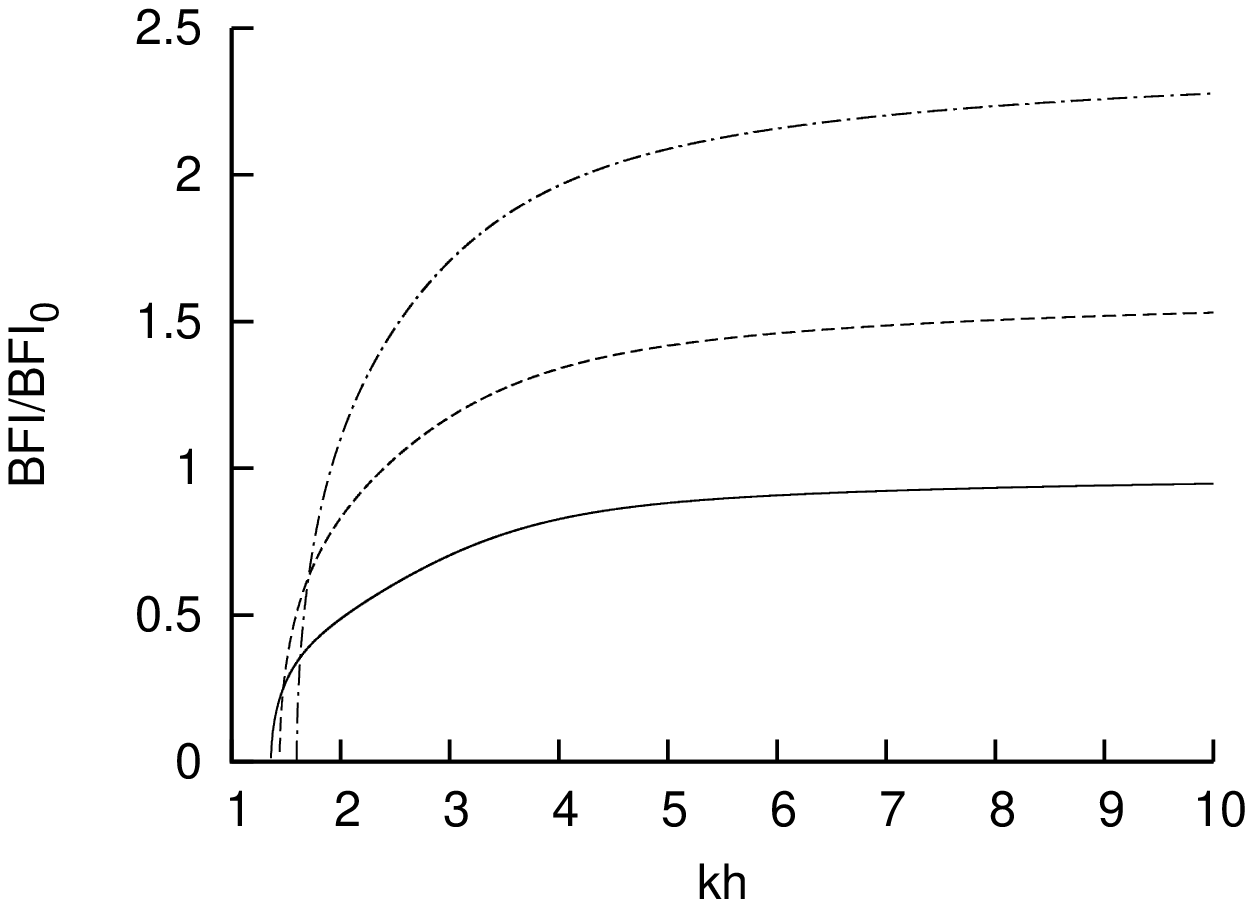}
\caption{Normalized Benjamin Feir Index as a function of $kh$ for several 
values of $\oOmega$~: $\oOmega=0.0$ (solid line), $\oOmega=1.0$ (dashed line),
$\oOmega=2.0$ (dot-dashed line)}
\label{BFI1}
\end{minipage} \hfill
\begin{minipage}[b]{.46\linewidth}
\includegraphics[width=\linewidth]{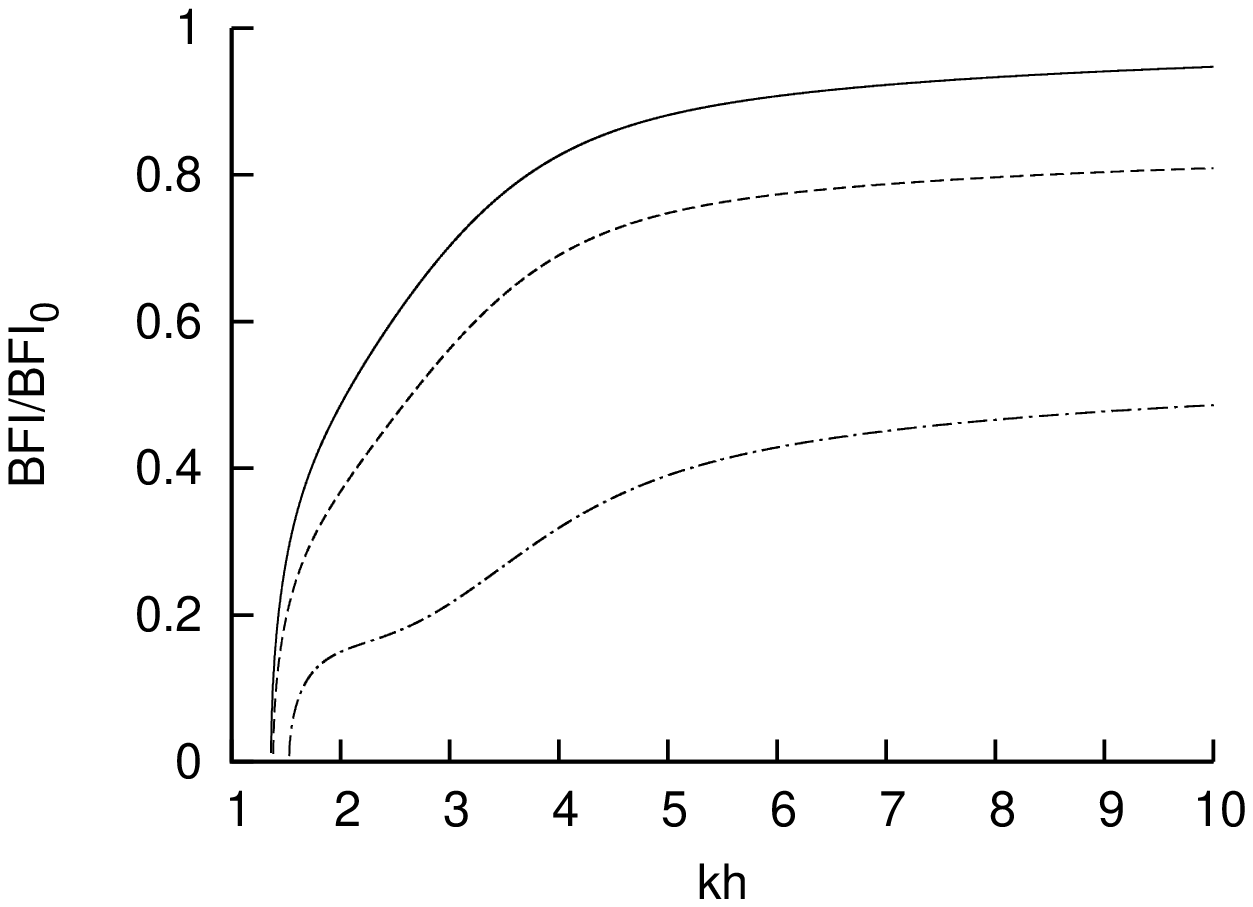}
\caption{Normalized Benjamin Feir Index as a function of $kh$ for several 
values of $\oOmega$~: $\oOmega=0.0$ (solid line), $\oOmega=-0.3$ (dashed line,
$\oOmega=-0.6$ (dot-dashed line)}
\label{BFI2}
\end{minipage}
\end{figure}

Within the framework of random waves Janssen \cite{janssen} introduced the concept of the Benjamin-Feir Index (BFI)
which is the ratio of the mean square slope to the normalized width of the spectrum. When this parameter
is larger than one, the random wave field is modulationally unstable, otherwise it is modulationally stable.
From the NLS equation Onorato {\em et al.} \cite{onorato} define the BFI as follows
\begin{equation}
BFI=\frac{a_0k}{\Delta k/k}\sqrt{\abs{ \frac{M_1}{L_1}}}
\end{equation}
where $\Delta k$ represents a typical spectral bandwidth.
\newline
In infinite depth the BFI without shear current is
\begin{equation}
BFI_0=\frac{4a_0k}{\Delta k/k}
\end{equation}
Hence, the normalized BFI writes
\begin{equation}
\frac{BFI}{BFI_0}= \frac{1}{4}\sqrt{\abs{\frac{M_1}{L_1}}}
\end{equation}
The coefficients $M_1$ and $L_1$ depend on the depth and vorticity. 
Onorato {\em et al.} \cite{onorato} considered the effect of the depth on the BFI. 
Herein, besides depth effect a particular attention is paid on the influence of the vorticity on the BFI. 
In order to measure the vorticity effect on the BFI, the ratio of the BFI in the presence of vorticity 
to its value in the absence of vorticity in infinite depth is plotted in figures \ref{BFI1} and \ref{BFI2}. 
For fixed value of $\oOmega$ the BFI increases with depth. Our results for $\oOmega=0$ 
are in full agreement with those of Onorato {\em et al.} \cite{onorato} (the solid line in figure \ref{BFI1}). 
Furthermore, it is shown for $\oOmega>0$ and sufficiently deep water that the BFI increases 
with the magnitude of the vorticity. Therefore, we may expect that the number of rogue waves 
increases in the presence of shear currents co-flowing with the waves. For $\oOmega<0$ the presence 
of vorticity decreases the BFI. 
For a more complete information about rogue waves, one may consult Kharif, Pelinovsky and Slunyaev (2009)\cite{KPS}.

\section{Conclusion}\label{conclusion}
Using the method of multiple scales, a $1D$ nonlinear Schr\"{o}dinger 
equation has been derived 
in the presence of a shear current of non zero constant vorticity in arbitrary depth. 
When the vorticity vanishes, the classical NLS equation is found. 
A stability analysis has been developed 
and the results agree with those of Oikawa {\em et al.} (1987) 
in the case of $1D$ NLS equation. 
We found that linear shear current may modify significantly 
the linear stability properties 
of weakly nonlinear Stokes waves.

\noindent We have shown the importance of the coupling between the mean flow
induced by the modulation and the vorticity. 
This coupling has been missed (or not emphasized) by previous authors.

\noindent Furthermore we have shown that the Benjamin-Feir instability 
can vanish in the presence of positive vorticity (\hbox{$\oOmega<0$}) for any depth.

\end{document}